\documentclass{aa}  
\usepackage{graphicx}
\usepackage{txfonts}
\usepackage{booktabs}

\begin{document}

   \title{Alignment in the orientation of LOFAR radio sources}


   \author{E. Osinga
          \inst{1}
          \and
          G. K. Miley\inst{1}
          \and 
          R. J. van Weeren\inst{1}
          \and 
          T. W. Shimwell\inst{1,2}
          \and 
          K. J. Duncan\inst{1,3}
          \and 
          M. J. Hardcastle\inst{4}
          \and
          A. P. Mechev\inst{1}
          \and
          H. J. A. R\"ottgering\inst{1}
          \and
          C. Tasse\inst{5,6}
          \and 
          W. L. Williams\inst{1}
          }

   \institute{Leiden Observatory, Leiden University, PO Box 9513, NL-2300 RA Leiden, The Netherlands \email{osinga@strw.leidenuniv.nl}
    \and
    ASTRON, the Netherlands Institute for Radio Astronomy, Postbus 2, NL-7990 AA Dwingeloo, The Netherlands
    \and
    Institute for Astronomy, Royal Observatory, Blackford Hill, Edinburgh, EH9 3HJ, UK 
    \and
    Centre for Astrophysics Research, University of Hertfordshire, College Lane, Hatfield AL10 9AB, UK
    \and 
    GEPI \& USN, Observatoire de Paris, Université PSL, CNRS, 5 Place Jules Janssen, 92190 Meudon, France
    \and 
    Department of Physics \& Electronics, Rhodes University, PO Box 94, Grahamstown, 6140, South Africa
    }

   \date{Received X; accepted Y}

 
  \abstract
    {Various studies have laid claim to finding an alignment of the polarization vectors or radio jets of active galactic nuclei (AGN) over large distances, but these results have proven controversial and so far, there is no clear explanation for this observed alignment. To investigate this case further, we tested the hypothesis that the position angles of radio galaxies are randomly oriented in the sky by using data from the Low-Frequency Array (LOFAR) Two-metre Sky Survey (LoTSS). A sample of 7,555 double-lobed radio galaxies was extracted from the list of 318,520 radio sources in the first data release of LoTSS at 150 MHz.
    We performed statistical tests for uniformity of the two-dimensional (2D) orientations for the complete 7,555 source sample. We also tested the orientation uniformity in three dimensions (3D) for the 4,212 source sub-sample with photometric or spectroscopic redshifts. Our sample shows a significant deviation from uniformity (p-value < $10^{-5}$) in the 2D analysis at angular scales of about four degrees, mainly caused by sources with the largest flux densities. No significant alignment was found in the 3D analysis. Although the 3D analysis has access to fewer sources and suffers from uncertainties in the photometric redshift, the lack of alignment in 3D points towards the cause of the observed effect being unknown systematics or biases that predominantly affect the brightest sources, although this has yet to be demonstrated irrefutably and should be the subject of subsequent studies.
    }


   \maketitle
%

\section{Introduction}\label{sec:introduction}

    The large sizes (up to few megaparsecs) of extended extragalactic radio sources allow us to use them in tracing the history of galactic nuclear activity over hundreds of millions of years. Since their discovery, it has been revealed that most powerful radio jets have highly linear morphologies \citep[e.g.,][]{1980ARA&A..18..165M}. In classical models of radio jets, the orientation is associated with the spin axis of a supermassive black hole (SMBH) in the nucleus of the host galaxy. The alignment of kpc and Mpc-scale radio emission with pc-scale jets \citep[e.g.,][]{Fomalont1975} has demonstrated that the collimated jets hold a "memory" of their directions for more than $10^8$ years. Our understanding of the accretion processes by which the SMBHs are "fed" or the mechanisms that determine the orientation of their spin axes is still incomplete. 
    
    An intriguing question concerns whether there could be some connection between the orientations of the SMBH spin axes and properties of the cosmic filaments in which the radio sources and their host galaxies are found. The possibility of such a connection has been suggested 
    in recent evidence for non-uniformity in radio-source position angles over large regions of the sky found by \citet{2016MNRAS.459L..36T} and \citet{2017MNRAS.472..636C}.

    If the radio sources are indeed aligned with respect to the large-scale structure in which they are found, a possible cause could be attributed to angular momentum transfer during the early stages of galaxy formation. The tidal torques imparted on the collapsing halos is found to influence the spin and shape of galaxies in N-body simulations \citep[e.g.,][]{1984ApJ...286...38W,2012MNRAS.427.3320C,2015MNRAS.446.2744L,2018MNRAS.481.4753C,2019arXiv190601623K}. However, the angular momentum vector of the active galactic nucleus (AGN) and the host galaxy are found to be misaligned and generally uncorrelated \citep{2012MNRAS.425.1121H}, indicating that this explanation is incorrect or incomplete.

    Furthermore, there is substantial evidence that  AGNs are associated with mergers, based on both observations and simulations \citep[e.g.,][]{2015ApJ...806..147C,2006MNRAS.365...11C}. If these mergers occur preferentially along the filaments of the large-scale structure, these could orient the central SMBHs in a particular way, resulting in a preferential alignment of the extended radio sources. Hence, if the alignment of radio sources on large scales is confirmed, this would have significant implications for models of the formation of galaxies and active galactic nuclei.
    
    Additional evidence that there may be a connection between the orientation of the spin axes of SMBHs that power active galactic nuclei and the cosmic filaments in which they lie comes from observations of large-scale statistical alignments in the optical polarization position angles of quasars \citep[e.g.,][]{1998A&A...332..410H,2001A&A...367..381H,2004MNRAS.347..394J}. Evidence has also been found for the polarization angle of quasars to be either parallel or perpendicular to the large-scale structures they inhabit \citep[e.g.,][]{2014A&A...572A..18H, 2016A&A...590A..53P}.
    
    A more extensive investigation of the large-scale distribution of radio source orientations is warranted. Surveys with the Low-Frequency Array (LOFAR) High Band Array \citep[HBA;][]{2013A&A...556A...2V} are especially suited for carrying out such studies, because they (i) are conducted at sufficiently low frequencies to detect steep-spectrum extended synchrotron radio structures, (ii) have sufficient angular resolution, with a $\sim$6$^{\prime\prime}$ half-power beam width (HPBW), to resolve 50 (100) kpc-sized radio sources out to redshift $\sim$ 1 ($>$6), and (iii) have the sensitivity and dynamic range needed to detect and measure orientations for an unprecedented number of sources. 
    
    Here, we describe such an investigation using position angles of radio sources from the LOFAR Two-metre Sky Survey Data Release I \citep[LoTSS-DR1;][]{2019A&A...622A...1S}. We first describe the data in Section \ref{sec:data}. The criteria we used to select sources with well-defined position angles from the 318,520 radio sources from the survey are discussed in Section \ref{sec:Selection}.
    The statistical methods we used to explore non-uniformity in the source alignments are explained in Section \ref{sec:statistics}. Our results are given in  Section \ref{sec:results}, where we report evidence for non-uniformity in the source alignments. Finally, in Sections \ref{sec:discussion} and \ref{sec:conclusion}, we discuss the robustness and implications of the results. 

    Throughout this paper, we adopt the Planck 15 cosmology \citep{2016A&A...594A..13P}. This cosmology is defined by the following relevant parameters: $H_0 = 67.8$ kms$^{-1}$Mpc$^{-1}$, $\Omega_m = 0.308$, $\Omega_\Lambda = 0.692$.


\section{The data}\label{sec:data}
    Our sample is taken from the LoTSS, a sensitive low-frequency (120-168 MHz) survey that will ultimately cover the entire northern sky. The first data release comprises 2\% of the whole survey (424 square degrees) in the HETDEX Spring Field region \citep[right ascension 10h45m to 15h30m and declination 45$^\circ$ to 57$^\circ$;][]{2019A&A...622A...1S}.
    It contains more than 300,000 radio sources that have a signal to noise ratio (S/N) of  $>$ 5. The images have a HPBW resolution of $\sim$ 6$^{\prime\prime}$, a median sensitivity of 71 $\mu$Jy/beam, and a positional accuracy better than $\sim$ 0.2$^{\prime\prime}$ 
   
   The data used were taken from the "value-added" radio + optical catalog of \citet{2019A&A...622A...2W} of 318,520 LoTSS sources, which includes, where possible, identifications and redshifts of the optical counterparts. The optical identifications were made using either a likelihood ratio method or by human visual classification through the LOFAR Galaxy Zoo\footnote{\url{https://www.zooniverse.org}}.  Spectroscopic redshifts in the added-value catalog were taken, where available, from the Sloan Digital Sky Survey Data Release 14 \citep{2018ApJS..235...42A}. Otherwise photometric redshifts were estimated using a hybrid methodology based on traditional template fitting and machine learning \citep[see][]{2019A&A...622A...3D}. 
   
\section{Source selection}\label{sec:Selection}

    For the alignment uniformity analysis, our goal was to select double-lobed radio sources with clearly defined position angles from the LoTSS value-added catalog. To identify such sources, we used the following method:

First, we filter the catalog to contain only high S/N extended sources. We define sources as extended if they have a major axis that is larger than five times the restoring beam size. The adopted selection criteria are: 
    \begin{equation*}
         S_{\mathrm{peak}}/N > 10 \textnormal{  and  } a > 30^{\prime\prime}, 
    \end{equation*}
where $S_{\mathrm{peak}}$ is the peak flux density of the LOFAR source at 144 MHz and $a$ is the size of the major axis of the source. The major and minor axes of some sources are not directly provided for sources that have been processed by the LOFAR Galaxy Zoo (LGZ). Instead, an equivalent "LGZ\_Size" and "LGZ\_Width" parameter is provided. The construction of the source dimensions from the LGZ data is described in \citet{2019A&A...622A...2W}. 
    Throughout this paper, we set the the major and minor axes of the sources processed by LGZ as the "LGZ\_Size" and "LGZ\_Width," respectively. Additionally, uncertainties for the LGZ shape parameters (source size, width, and position angle) are not provided by the value-added catalog. We discuss any effects due to uncertainties in the position angles in Section \ref{sec:discussion}.
    
    Next, we keep only the sources with a double lobed structure. We enforce this criterion by imposing the condition that sources must be fitted by multiple Gaussian components by the initial source finder PyBDSF  \citep{2015ascl.soft02007M}. This is indicated by the "S\_Code" of the source in the catalog. It is also possible that the source is a bright resolved nearby galaxy and these are identified with the "ID\_flag" code where the first digit is 2. We remove these sources as well, using:    \begin{equation*}
        ``S\_Code = M" \textnormal{  and  } ``ID\_flag" \neq 2.
    \end{equation*}
    
    Imposing these criteria results in a reduction of the sample of 318,520 sources to a sample of 7,688 bright extended linear sources. We check the catalog for sources that might have been identified multiple times by examining the distance from every source to its nearest neighbor. We investigate all sources that have a nearest neighbor within ten synthesized beams (1 arcminute). If the source has a different optical identification from its nearest neighbor, we can be reasonably sure that it is not a duplicate entry. When the source has an optical identification while the nearest neighbor does not, or both the source and the nearest neighbor lack an optical identification, we cannot be certain that these entries are not duplicates. To err on the side of caution, we remove all sources from our sample that have a nearest neighbor within ten synthesized beams, unless they have a different optical identification from their nearest neighbor. We find that 165 sources have a nearest neighbor within ten synthesized beams, and 32 of these have a different optical identification from their nearest neighbor. 
    We expect that removing the other 133 entries would not impact the strength of a possible alignment effect since radio source alignments have been claimed on scales of at least a degree \citep{2016MNRAS.459L..36T,2017MNRAS.472..636C} and these source separations are on a smaller angular scale than this. Thus, the final sample contains 7,555 selected sources.

\section{Statistical methods}\label{sec:statistics}
    To determine the departure from uniformity of the alignment of radio sources on the sky, an appropriate statistical method must be used that accounts for effects due to the geometry of the celestial sphere. We shall do this by introducing the concepts of "parallel transport" and "dispersion measure."
   
    \subsection{Parallel transport}
    The position angle in the LoTSS catalog is defined as the angle of the major axis of a source measured east of the local $m$ (north) direction. To have a consistent definition of the position angle across all pointings, we translated the position angles to be measured east of the direction of the north celestial pole.
    
    Because the position angle is defined with respect to the local meridian, the vectors corresponding to the position angles on different points of the celestial sphere cannot be compared directly. These vectors must be transported along the great circle joining these points. Following \citet{2004MNRAS.347..394J} and \citet{2017MNRAS.472..636C}, we use the parallel transport method, by which the radio source "vectors" can be transported to a different position on the celestial sphere. This method is described below for completeness.
    
    \begin{figure}[t] 
        \centering
        \includegraphics[width=\columnwidth]{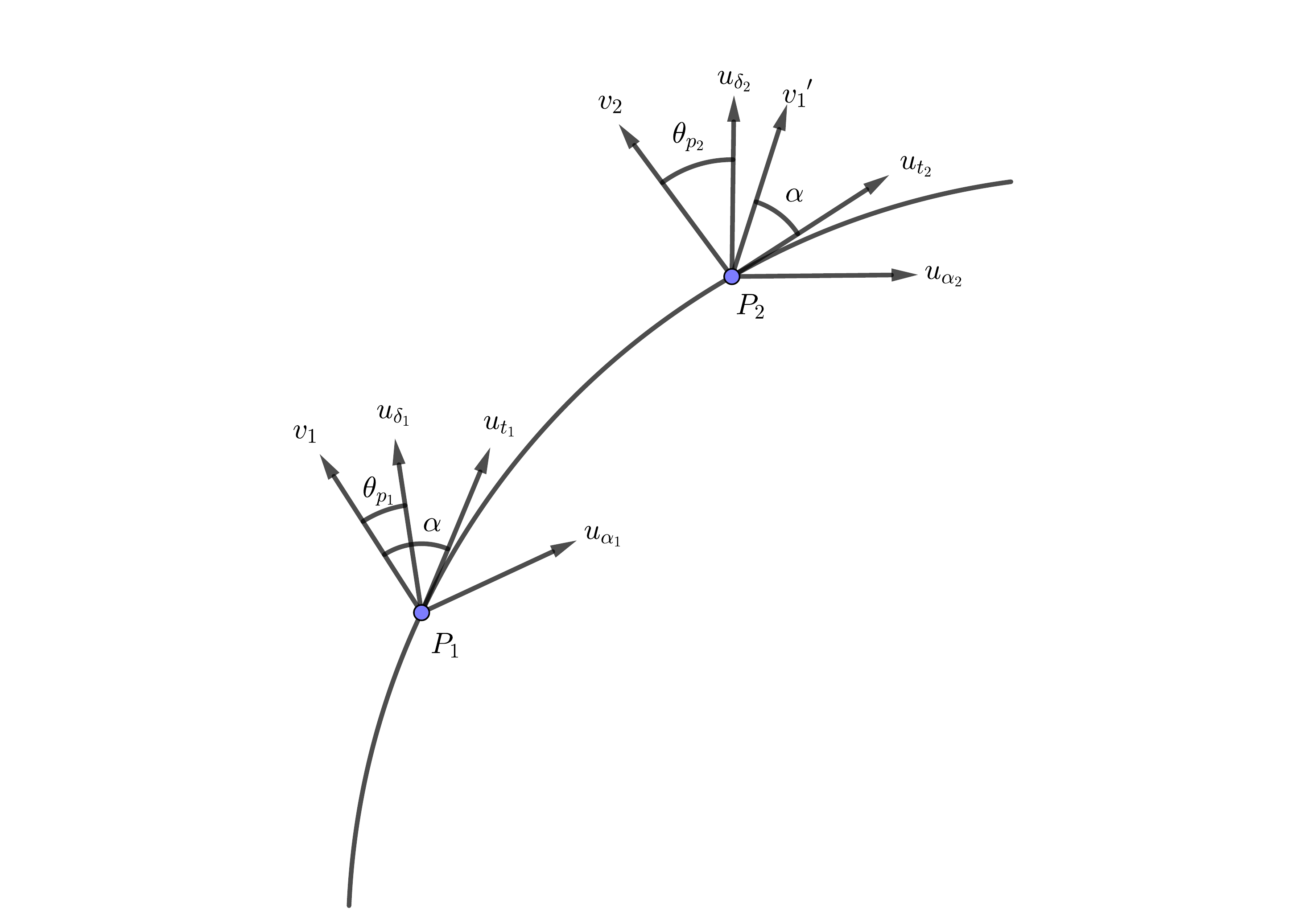}
        \caption{Illustration of parallel transport. Vector $v_1$ corresponding to a position angle $\theta_{p_1}$ and vector $v_2$ corresponding to position angle $\theta_{p_2}$ are shown. In order to compare $v_1$ to $v_2$, $v_1$ must be parallel-transported along the great circle indicated by the curve from location $P_1$ to location $P_2$. The transported vector is indicated by $v_1'$ and the local basis vectors are denoted by ($u_\delta$,$u_\alpha$). In parallel transport, the angle $\alpha$ between the vector tangent to the sphere $u_{t}$ and the vector $v$ remains fixed. Figure adapted from \citet{2004MNRAS.347..394J}.}
        \label{fig:parallel_transport}
    \end{figure}

    We parametrize the celestial sphere with local unit vectors $(\vec{u}_r, \vec{u}_\delta, \vec{u}_\alpha)$ which point, respectively, to the center of the sphere, north along the local meridian and eastwards on the sphere. We wish to compare the position angles, $\theta_{p_1}$ and $\theta_{p_2}$ , of sources 1 and 2 with positions, $P_1$ and $P_2,$ on the celestial sphere (Figure \ref{fig:parallel_transport}). The vector resulting from the position angle, $\theta_{p_1}$ , of source $1$ at location, $P_1,$ is given, in terms of the local basis, by
    
    \begin{equation}
    \vec{v}_1 = \cos\theta_{p_1} \vec{u}_{\delta_1} + \sin\theta_{p_1} \vec{u}_{\alpha_1}.
    \end{equation}
    To define a coordinate-invariant inner product we parallel transport the vector, $\vec{v}_1,$ to the position, $P_2,$ to obtain vector, $\vec{v_1}'$. Vector $\vec{v_1}'$ then makes an angle, $\theta_p '$ , with respect to the local north-pointing vector, $\vec{u}_{\delta_2}$. To find the transported angle ,$\theta_p '$, let $\vec{u}_s$ be the unit vector perpendicular to the plane containing the two radial vectors $\vec{u}_{r_1}$ and $\vec{u}_{r_2}$. Thus, $\vec{u_s}$ is found by
    
    \begin{equation}
    \vec{u}_s = \frac{\vec{u}_{r_1} \times \vec{u}_{r_2}}{ {|\vec{u}_{r_1}} \times \vec{u}_{r_2}| }.
    \end{equation}
    Consider now the unit vectors, $\vec{u}_{t_1}$ and $\vec{u}_{t_2}$ , at points, $P_1$ and $P_2$, and tangent to the great circle connecting $P_1$ and $P_2$. These vectors are given by:
    
    \begin{equation}
    \vec{u}_{t_{1,2}} = \vec{u}_{s} \times \vec{u}_{r_{1,2}}.
    \end{equation}
    In terms of the local basis, these vectors can be written as:
    
    \begin{equation}
    \vec{u}_{t_1} = \vec{u}_{\delta_1} \cdot \vec{u}_{t_1}\vec{u}_{\delta_1} + \vec{u}_{\alpha_1} \cdot \vec{u}_{t_1}\vec{u}_{\alpha_1},
    \end{equation}
    where
    \begin{equation}
    \vec{u}_{\delta_1} \cdot \vec{u}_{t_1} = \frac{ -\sin\delta_1\cos\delta_2 + \cos\delta_1\sin\delta_2\cos(\theta_{p_1} - \theta_{p_2} ) }{\sqrt[]{1-\left(\vec{u}_{r_1}\cdot \vec{u}_{r_2}\right)^2}},
    \end{equation}
    
    \begin{equation}
    \vec{u}_{\alpha_1} \cdot \vec{u}_{t_1} = \frac{ \sin\delta_2\sin(\alpha_2 - \alpha_1)}
    {\sqrt[]{1-\left(\vec{u}_{r_1}\cdot \vec{u}_{r_2}\right)^2}},
    \end{equation}
    
    \begin{equation}
    \vec{u}_{\delta_2} \cdot \vec{u}_{t_2} = \frac{ -\sin\delta_2\cos\delta_1 + \cos\delta_2\sin\delta_1\cos(\theta_{p_1} - \theta_{p_2} ) }{\sqrt[]{1-\left(\vec{u}_{r_1}\cdot \vec{u}_{r_2}\right)^2}},
    \end{equation}
    
    \begin{equation}
    \vec{u}_{\alpha,2} \cdot \vec{u}_{t_2} = \frac{ -\sin\delta_1\sin(\alpha_1 - \alpha_2)}
    {\sqrt[]{1-\left(\vec{u}_{r_1}\cdot \vec{u}_{r_2}\right)^2}}.
    \end{equation}

    As $\vec{v}_1$ is parallel-transported along the great circle to position, $P_2$ , with its angle with respect to the tangent of the great circle remaining fixed. Thus, to determine the angle by which the vector has turned due to this transport, we consider the orientation of $\vec{u}_{t_1}$ and $\vec{u}_{t_2}$ with respect to the local basis at the two points where the sources lie. We call $\xi_1$ the angle between $\vec{u}_{t_1}$ and $\vec{u}_{\alpha_1}$ , and $\xi_2$ the angle between $\vec{u}_{t_2}$ and $\vec{u}_{\alpha_2}$. These angles are given, per definition of the inner product, by
    
    \begin{equation}
    \xi_{1,2} = \arccos(\vec{u}_{\alpha_{1,2}} \cdot \vec{u}_{t_{1,2}})
    .\end{equation}
    Thus, the transported $\vec{v_1}'$  makes an angle $\theta_{p,1}' = \theta_{p,1} + (\xi_2 - \xi_1)$ defined with respect to the local coordinates in $P_2$. Hence we can now define the generalized dot product between $\vec{v}_1$ and $\vec{v}_2$ as the dot product between the transported vector $\vec{v_1}'$ and $\vec{v}_2$:
    
    \begin{equation}\label{eq:old_dotproduct}
    \vec{v}_1 \odot \vec{v}_2 = \vec{v_1}' \cdot \vec{v}_2 = \cos(\theta_{p_1} - \theta_{p_2} + \xi_2 - \xi_1).
    \end{equation}
    
    Equation \ref{eq:old_dotproduct} can generally be used in any problem that considers angles on a sphere. In particular, when comparing the difference between position angles, it makes sense to redefine the generalized inner product between two position angles as 
    
    \begin{equation}\label{eq:dotproduct}
    (\theta_{p_1},\theta_{p_2}) = \cos[2(\theta_{p_1}-\theta_{p_2} + \xi_2 - \xi_1)],
    \end{equation}
    where, since the position angles range from 0 to $\pi$, it assumes values of $\in$ $(-1,1)$ and where $+1$ expresses the perfect alignment between $\theta_{p_1}$ and $\theta_{p_2}$ and $-1$ indicates perpendicular orientations.
    
    \subsection{Statistical test}\label{sect:statisticaltest}
    
    To test the significance of a possible alignment in source position angles, we use the dispersion measure \citep{2004MNRAS.347..394J,2017MNRAS.472..636C}. We briefly repeat the definition of the dispersion measure here for completeness.
    
    The dispersion depends only on the differences between neighboring position angles and it is, therefore, a suitable choice when testing for alignment on different scales. The dispersion measure of source, $i,$ as a function of a position angle, $\theta,$ is defined as 
    
    \begin{equation}\label{eq:dispersion}
    d_{i,n}(\theta) = \frac{1}{n} \sum_{k=1}^{n}(\theta,\theta_k),
    \end{equation}
    where $n$ is the number of nearest neighbors that are considered around source $i$, including the source itself, and $\theta_k$ is the position angle of the respective neighbors. The generalized inner product $(\theta,\theta_k)$ is defined by Equation \ref{eq:dotproduct}. 
    
    The position angle $\theta$ that maximizes the dispersion around source $i$ is analogous to the definition of the mean position angle of source $i$ and its $n$ nearest neighbor s. The magnitude of $d_{i,n}\vert_{max}$ is, then, a measure of the dispersion around this mean. The dispersion can take a maximum value of 1, which corresponds to perfect alignment of all $n$ nearest neighbors. To find the value of $\theta$ that maximizes the dispersion, we take the derivative of Equation \ref{eq:dispersion} with respect to $\theta$ and, after some intermediate steps, we arrive at the following expression for $d_{i,n}\vert_{max}$:
    
    \begin{equation}\label{eq:dispersionmax}
    d_{i,n}\vert_{max} = \frac{1}{n} \left[ \left(\sum_{k=1}^{n} \cos \theta_k\right)^2 + \left(\sum_{k=1}^{n} \sin \theta_k\right)^2 \right]^{1/2}.
    \end{equation}
    The statistic, so that we may test for the non-uniformity of alignment in a sample of $N$ sources, is then defined as: 
    
    \begin{equation}\label{eq:statisticS}
    S_{n} = \frac{1}{N} \sum_{i=1}^{N} d_{i,n}\vert_{max},
    \end{equation}
    which is simply the average of the maximum dispersion for a number of nearest neighbors, $n,$ calculated over all $N$ sources in the sample. This statistic thus measures the strength of a local alignment signal in the full sample of $N$ sources while considering the $n$ nearest neighbors of every source. 
    
    The significance level for rejecting the null hypothesis that a sample of sources is randomly oriented is then given by comparing the statistic of the dataset, $S_{n}$ , to the distribution of the statistic for  simulated  samples that are randomly oriented. It is found through a one-tailed significance test, expressed as:
    \begin{equation}\label{eq:SignificanceLevel}
    SL = 1 - \Phi\left( \frac{S_{n} - \left\langle S_{n}\vert_{MC}\right\rangle}{\sigma_n}\right),
    \end{equation}
    where $\Phi$ is the cumulative standard normal distribution function. Here, $<S_{n}\vert_{MC}>$ and $\sigma_n$ are, respectively, the expectation value and standard deviation of $S_{n}$ in the absence of alignment. These values can be found through Monte Carlo simulations of randomly oriented sources.
    
    \citet{2004MNRAS.347..394J} verified that for randomly oriented samples of sources, $S_{n}$ is normally distributed if $N\gg{}n\gg{}1$ is satisfied.
    With the dispersion measure and the resulting statistic, the significance level at which the hypothesis of uniformity in the position angles should be rejected can be calculated on a local scale by probing different numbers of nearest neighbors. Since the number of nearest neighbors can be translated to fixing apertures with angular radii extending to the $n$-th nearest neighbor around all sources, $S_{n}$ can be used to probe the significance of alignment on different angular scales. We note that different $S_{n}$ are not independent since the dispersion is an average of $n$ neighbors. This statistic thus probes alignment up to scales corresponding to $n$ and once a signal is detected for some $n$, a preferentially positive signal is expected for larger $n$.
    
    If the redshifts of the sources are known, this method can be extended to probing nearest neighbors in 3D space. In this way, the dependence of a possible alignment effect and $S_{n}$ as a function of physical scale can be probed.

\section{Results}\label{sec:results}

\begin{figure}[t] 
      \centering
      \includegraphics[width=\columnwidth]{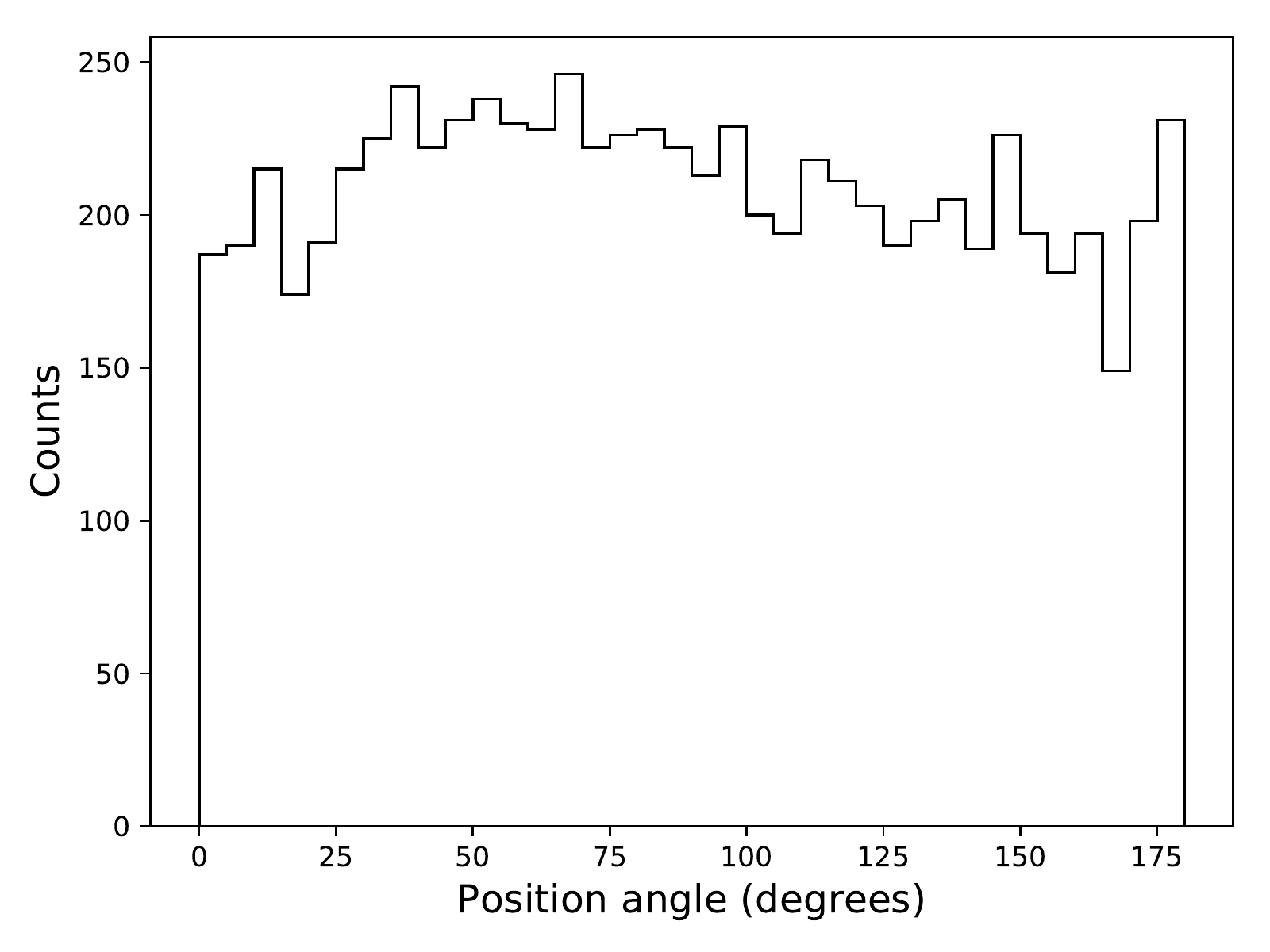}
      \caption{Position angle distribution of the complete sample of selected sources.}
      \label{fig:dist}
    \end{figure}

    We first tested the uniformity of the LoTSS radio source position angles over the complete 424 square degrees of the available survey to give an indication of possible systematic effects. The distribution of position angles is given in Figure \ref{fig:dist}. We expect the position angles to be uniformly distributed over this relatively large patch of the sky, if no systematic effects are present. From Figure \ref{fig:dist}, we can see that no major systematic effects are present, although the distribution is not quite uniform. To check if the distribution is consistent with a uniform distribution of sources, we applied the Kolmogorov-Smirnov (K-S) test \citep[e.g.,][]{2012JCAP...01..009F}. The K-S test resulted in a p-value of 0.030 per cent. This is strong evidence for rejecting the null hypothesis that the distribution of position angles over the complete sample is uniform, which indicates some systematic (survey-wide) bias in our sample. Still, the local alignment signal might be stronger or weaker depending on the nature of the effect that is causing the alignment.

    \subsection{Two-dimensional analysis}\label{sect:2D}
    To determine whether the hypothesis of uniformity in position angles on different angular scales should be rejected and if so, at what significance level, we compared results for the observed LoTSS sample with those for 1,000 simulated randomly distributed position angle samples. These samples were generated by randomly shuffling the position angles among the sources to maintain the same global position angle distribution and source positions.

\begin{figure}[t] 
    \centering
    \includegraphics[width=1.0\columnwidth]{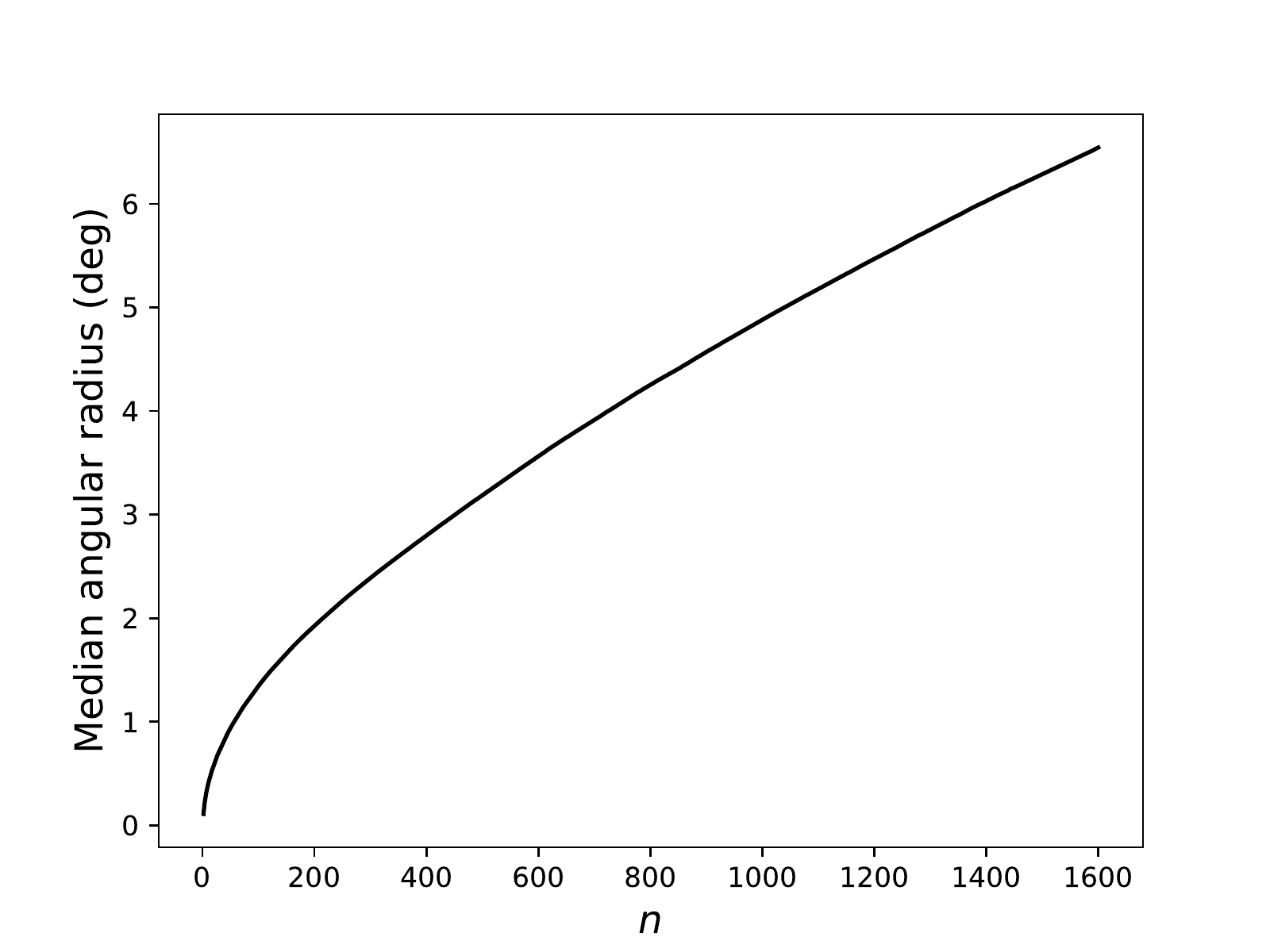}
      \caption{Median angular radius corresponding to drawing a circular aperture around every source with an angular radius bound by the $n$-th neighbor around that source.}
      \label{fig:angular_radius}
    \end{figure}

    The sample was checked for local alignment by probing the statistic, $S_{n}$ , for different numbers of nearest neighbors. To express the statistic in terms of angular scale, a circular aperture with a radius extending to the $n$-th neighbor of every source is drawn. We translated the number of nearest neighbors to an approximate corresponding angular scale by taking the median angular radius of all these apertures. This dependency is shown in Figure \ref{fig:angular_radius}. 
    
\begin{figure}[t]
      \centering
      \includegraphics[width=1.0\columnwidth]{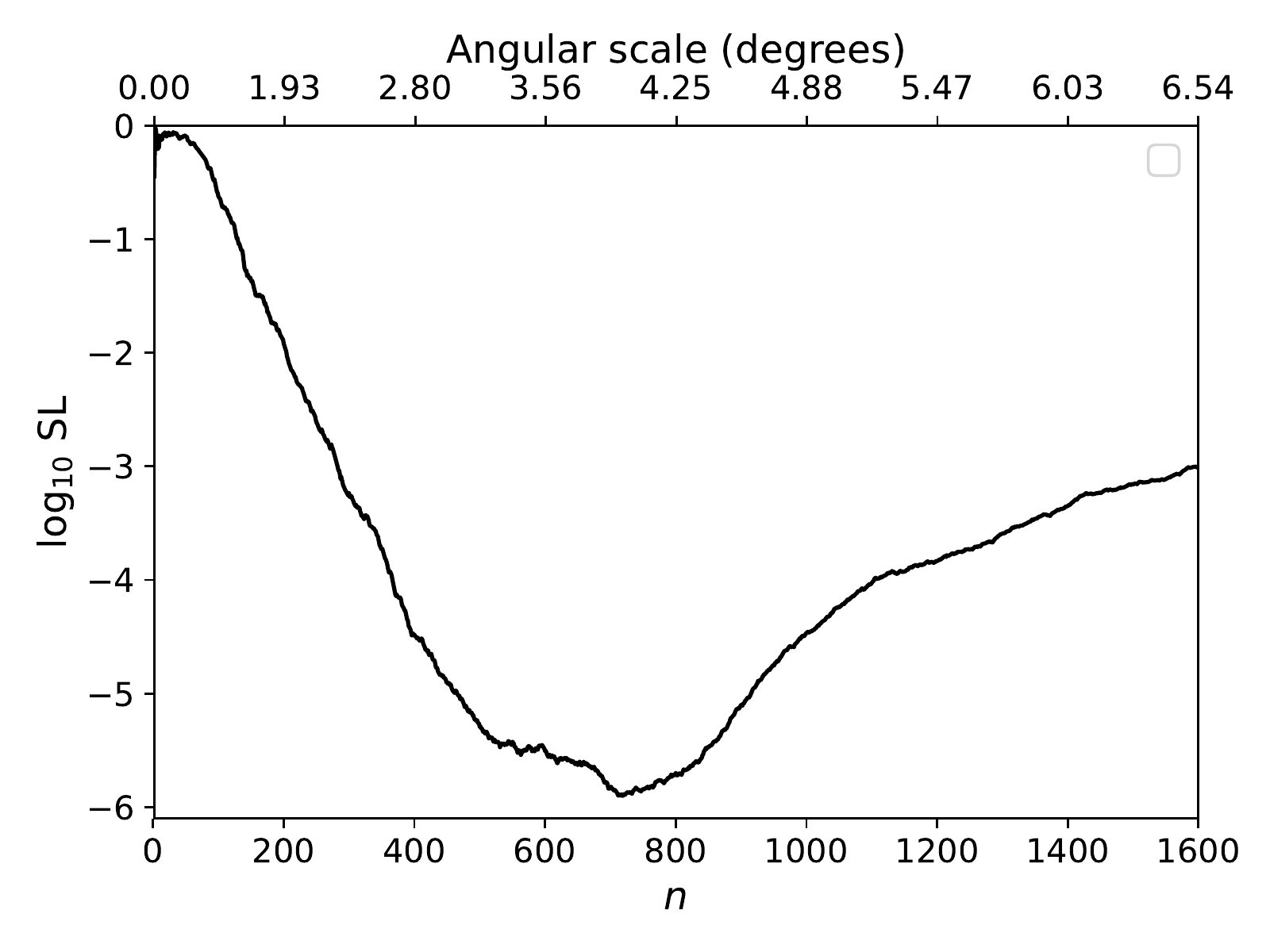}
      \caption{Logarithm of the significance level for which uniformity in position angles as a function of the number of nearest neighbors $n$ should be rejected for the sample of 7,555 selected sources. The conversion to angular scale is shown in Fig. \ref{fig:angular_radius}.}
      \label{fig:SL_allsubsample} 
    \end{figure}

    The significance level at which the null hypothesis should be rejected (of the position angles being uniformly distributed) is given as a function of the number of nearest neighbors (or corresponding angular scale) in Figure \ref{fig:SL_allsubsample}. There is strong evidence that the hypothesis of uniformity in radio source position angles should be rejected on angular scales of about four degrees, with a significance level of < $10^{-5}$.

    \begin{table*}[]
    \centering
    \caption{Parameters that cut the initial sample of selected sources into four equal frequency total flux density $f$ bins. The maximum significance level to reject uniformity is also shown.}
    \label{tab:fluxbins}
    \begin{tabular}{@{}llllll@{}}
    \toprule
    Bin number & Flux range (mJy) & Median flux (mJy) & Median redshift & Median size ($^{\prime\prime}$) & Significance level \\ \midrule
    0 & $f$ \textless $12$ & 7 & 0.55 & 42 & 1.1$\cdot10^{-2}$ \\
    1 & $12$ \textless $f$ \textless $33$ & 20 & 0.54 & 51 & 1.5$\cdot10^{-1}$ \\
    2 & $33$ \textless $f$ \textless $96$ & 54 & 0.57 & 59 & 2.9$\cdot10^{-1}$ \\
    3 & $96$ \textless $f$ & 227 & 0.63 & 68 & 7.7$\cdot10^{-11}$ \\ \bottomrule
    \end{tabular}%
    \end{table*}

\begin{figure}[] 
        \centering
        \includegraphics[width=\columnwidth]{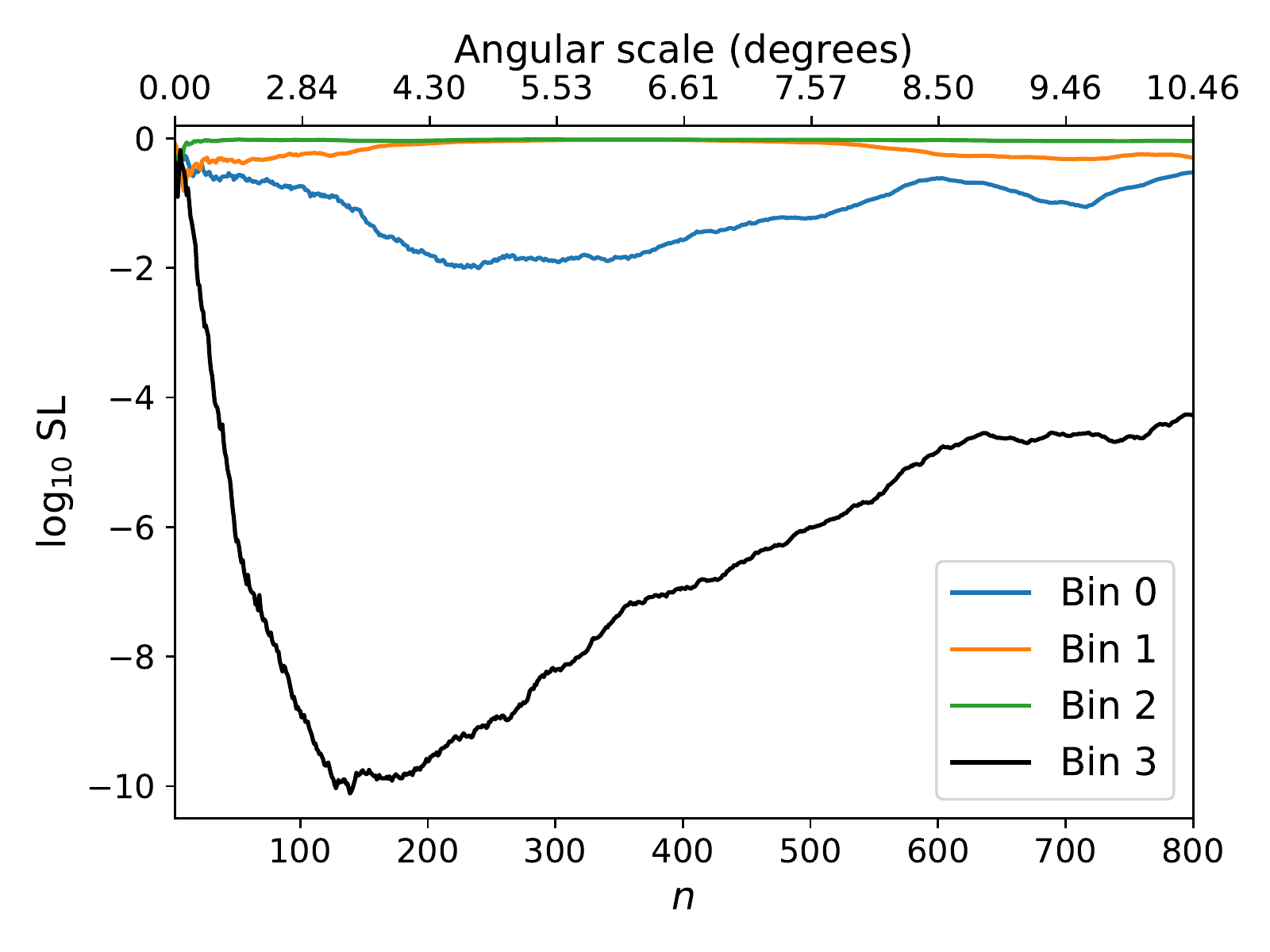}
        \caption{Logarithm of the significance level for the sample of sources split into four equal frequency total flux density bins, as defined in Table \ref{tab:fluxbins}.}
        \label{fig:fluxbins2D}
    \end{figure}

    To investigate the effect further, we split our sample into four equal frequency flux density bins to have the maximum number of sources in every bin, as given in Table \ref{tab:fluxbins}. For each bin, this table includes the median flux density, the median redshift, the median source angular size, and the maximum significance level at which the null-hypothesis of position angle uniformity should be rejected, taken from Figure \ref{fig:fluxbins2D}.

    Figure \ref{fig:fluxbins2D} shows the significance level of position angle alignment for the four flux density bins as a function of angular distance. Interestingly, the highest flux density bin shows very strong evidence for alignment, up to scales of roughly ten degrees, but most significantly around four degrees, while all other bins are consistent with uniform distributions. This shows that the effect seen in the total sample is caused by the highest flux density sources only. 
    
    \subsection{Three-dimensional analysis} \label{sect:3D}
\begin{figure}[] 
        \centering
        \includegraphics[width=\columnwidth]{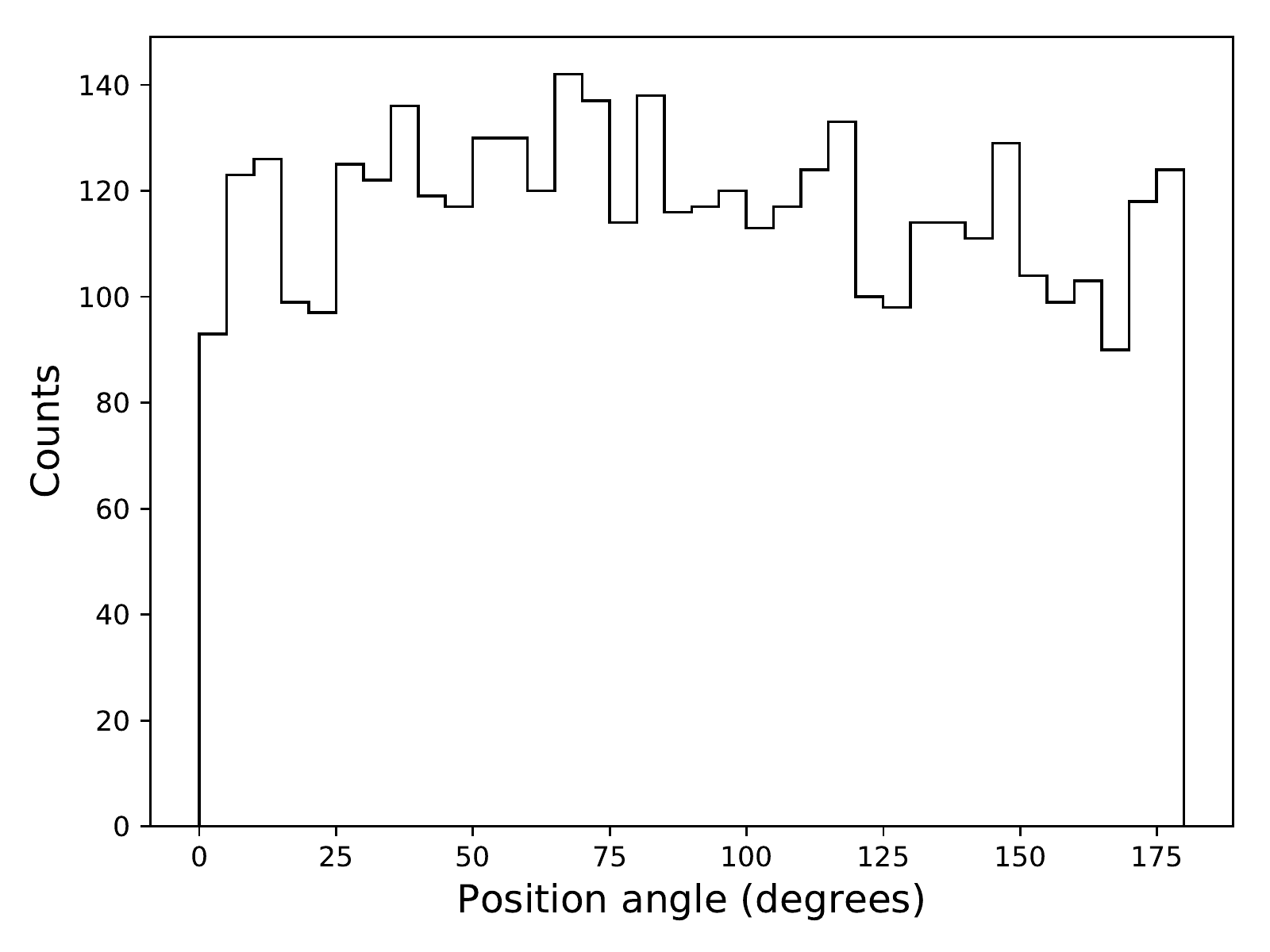}
        \caption{Distribution of position angles for the sample of 4,212 sources that have a redshift measurement.}
        \label{fig:hist_all_subsample3D}
    \end{figure}

    We carried out an analysis of alignment uniformity using 3D source positions, after removing all sources from our sample that do not have a spectroscopic or photometric redshift tabulated in the value-added catalog \citep{2019A&A...622A...2W,2019A&A...622A...3D}. This reduced the size of our sample to 4,212 sources. The number of photometric and spectroscopic redshifts is 2,311 and 1,901, respectively. We emphasize that the statistical method is exactly the same for this analysis. The only difference between the 2D and 3D analysis is that 3D source positions are now used to find the $n$ nearest neighbors for every source.
    The distribution of position angles of these 4,212 sources is shown in Figure \ref{fig:hist_all_subsample3D}. The K-S test indicates a p-value of 1.0 per cent, indicating that for this sample of 4,212 sources, there is weak evidence for rejecting the null hypothesis for the uniformity of position angles.

    In the analysis for local alignment, the samples were compared again with 1,000 simulated uniformly distributed position angle samples, generated by randomly shuffling the position angles among the sources. To repeat the analysis in three dimensions, each source was assigned a position in 3D space according to their right ascension, $\alpha$, declination, $\delta,$ and comoving distance, $r,$ as follows:
    
    \begin{equation}
    \begin{array}{l}
    x = r \cos \alpha \cos \delta,  \\
    y = r \sin \alpha \cos \delta, \\
    z = r \sin \delta. 
    \end{array}
    \end{equation}
    The nearest neighbors were then computed in 3D space according to these positions to probe for alignment on local scales. 
    
\begin{figure}[]{}
      \centering
      \includegraphics[width=\columnwidth]{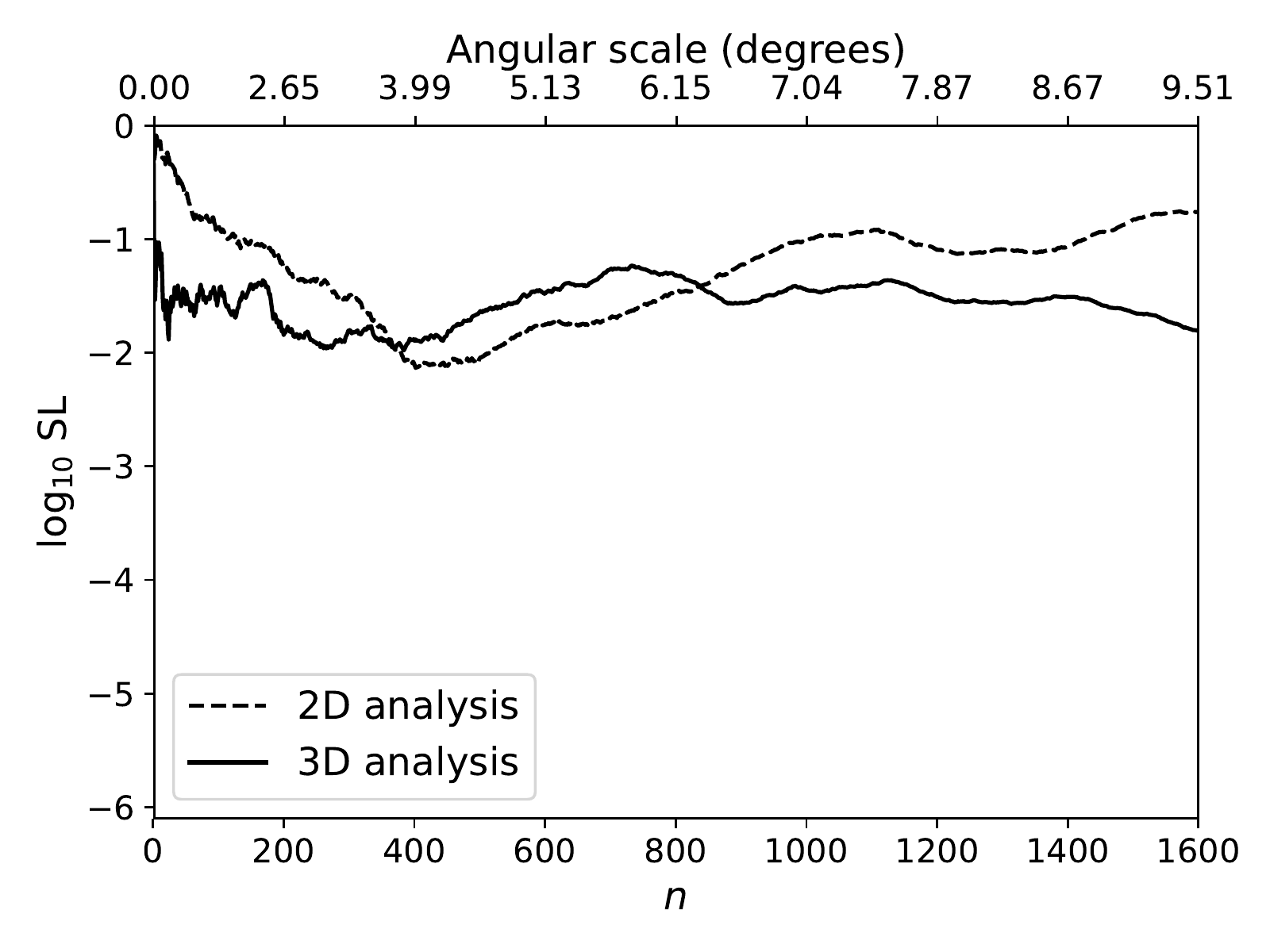}
      \caption{Logarithm of the significance level at which position angle uniformity should be rejected, as a function of the number of nearest neighbors $n$ for the 4,212 sources in that have redshifts available. The dashed line indicates the results of the 2D analysis and the solid line the results of the 3D analysis.}
      \label{fig:SL_allsubsample3D} 
    \end{figure}

    Figure \ref{fig:SL_allsubsample3D} shows the significance level at which the hypothesis of uniformity in position angles can be rejected for the 4,212 sources that have a redshift, both in a 3D and a 2D analysis. The 3D analysis does not show strong evidence for an alignment effect. 
    
    Since the sources with the largest flux densities are the main contributor to the alignment effect in the 2D analysis, we also calculated the significance for the highest flux density sources in the 3D analysis. We split the 4,212 sources into four equal frequency total flux density bins, which defines the highest flux density bin as all sources with a total flux density > $108$ mJy.
    This makes the flux cut for the highest flux density bin slightly higher than the equivalent in the 2D analysis, but we decide to use this flux cut to have a fairer comparison between the different flux density bins within the 3D analysis.
    
\begin{figure}
        \centering
        \includegraphics[width=\columnwidth]{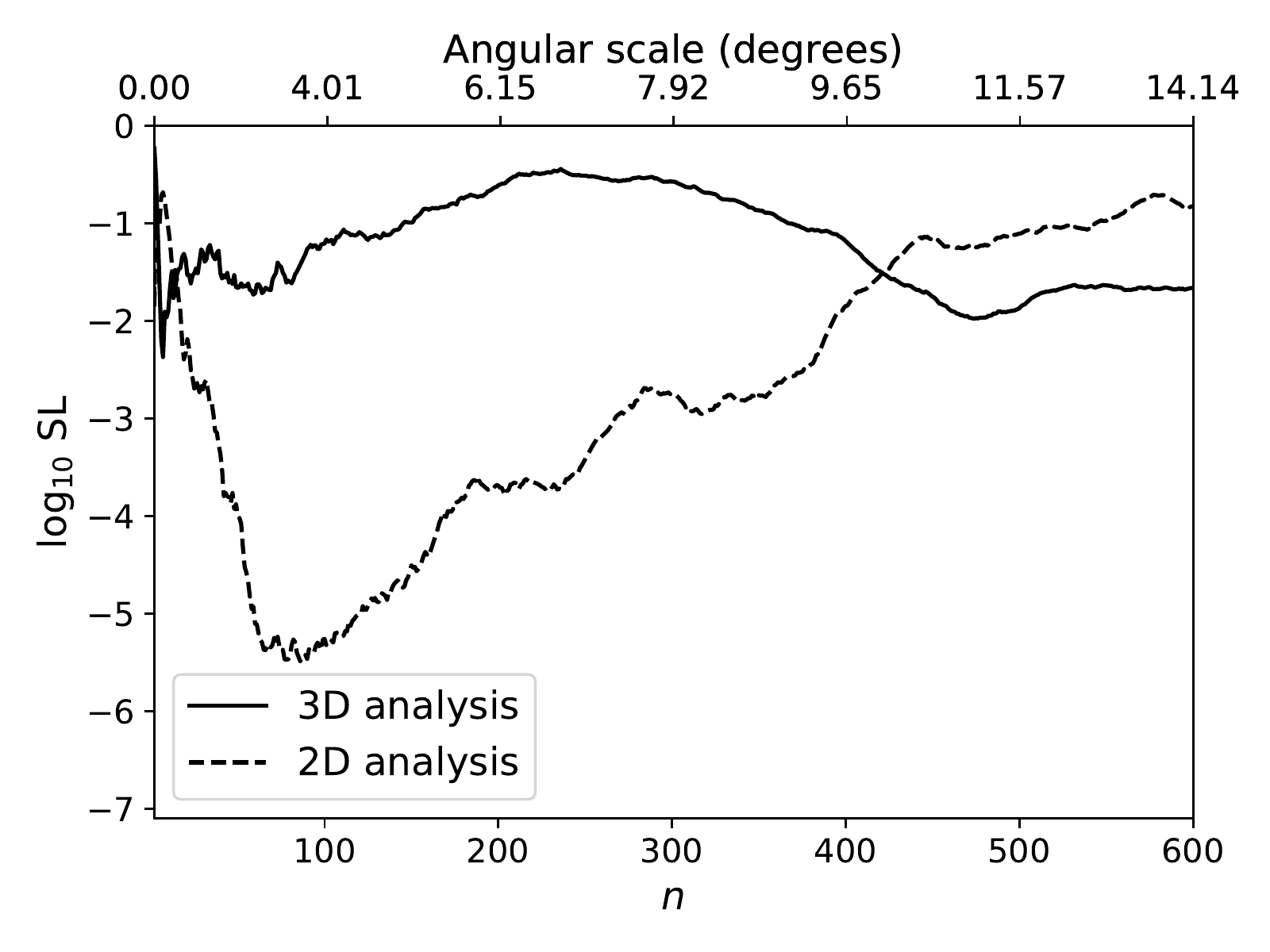}
        \caption{Logarithm of the significance level at which position angle uniformity should be rejected as a function of the number of nearest neighbors $n$ for the 1,051 sources with total flux density > 108 mJy and a redshift measurement. The dashed line indicates the results of the 2D analysis and the solid line the results of the 3D analysis.}
        \label{fig:SL_flux3_2D_3D}
    \end{figure}

    The significance level at which position angle uniformity can be rejected for the highest flux density bin in 3D is shown in Figure \ref{fig:SL_flux3_2D_3D}. This figure shows, interestingly, that the 2D analysis of these 1,051 sources still shows strong evidence for alignment up to scales of four degrees. However, this signal is not present in the 3D analysis. No signal was found in the other flux density bins, either in 2D or in 3D.
    
    The difference between the 2D and 3D analysis indicates that the 2D alignment effect is due to some unknown systematic effect, since a physical effect would invariably cause stronger alignment in the 3D analysis than in the 2D analysis. Additionally, we inspected whether the most radio luminous sources are also the most aligned sources, which would be expected from the similar median redshift per flux density bin. However, no alignment signal was found in either the 2D or 3D analysis of the 1,051 highest radio power sources.

    \begin{figure}
        \centering
        \includegraphics[width=\columnwidth]{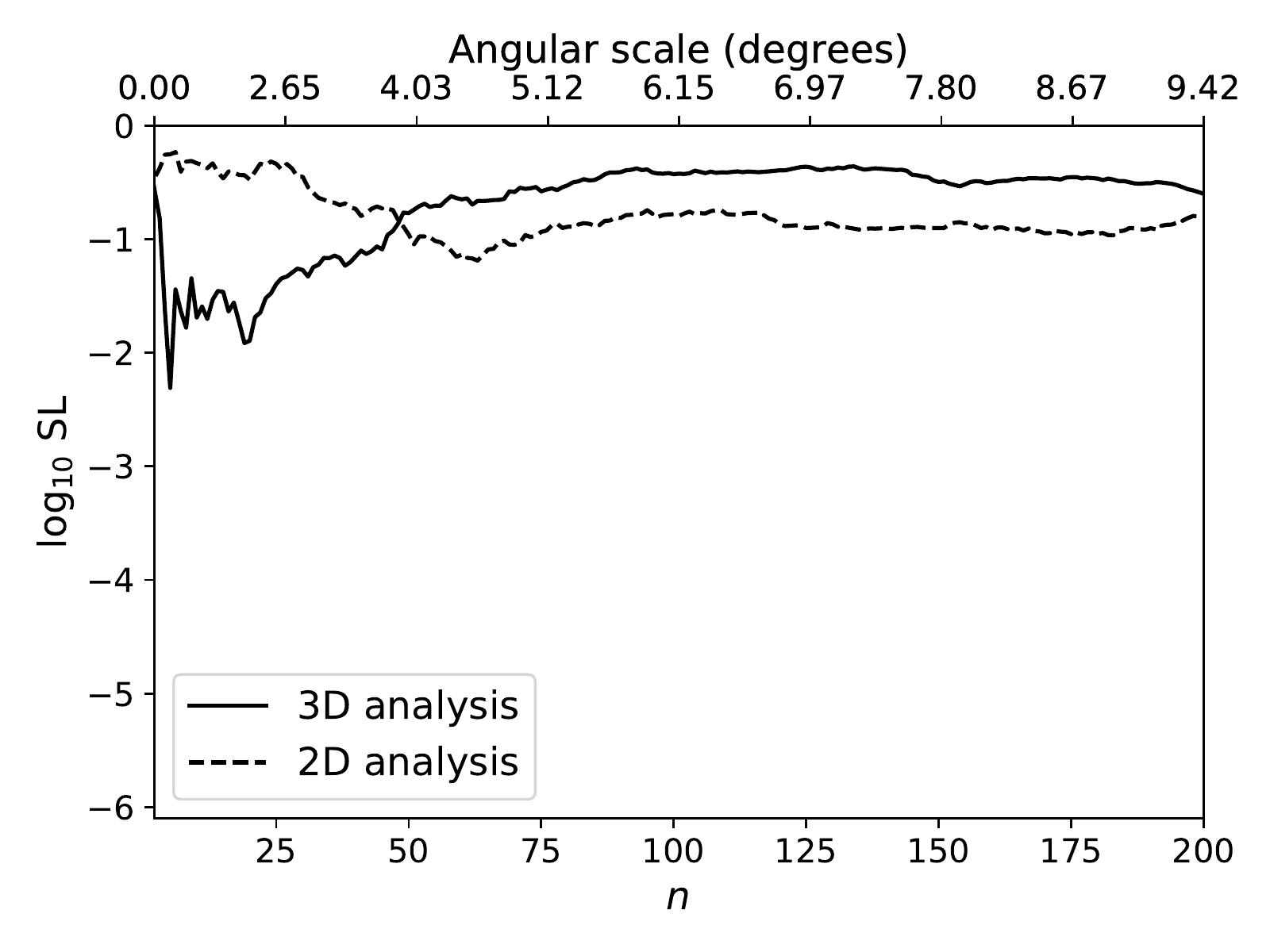}
        \caption{Logarithm of the significance level at which position angle uniformity should be rejected as a function of the number of nearest neighbors $n$ for the 523 sources with total flux density > 108 mJy and a spectroscopic redshift measurement. The dashed line indicates the results of the 2D analysis and the solid line the results of the 3D analysis.}
        \label{fig:spectroscopic_sources_only_fluxbins}
    \end{figure}
    
    Although it reduces the sub-sample sizes even further, we also tested if the results depend on whether the redshifts were photometric or spectroscopic. Figure \ref{fig:spectroscopic_sources_only_fluxbins} shows the results for the 523 sources that have a spectroscopic redshift. The figure shows that in both the 2D analysis and 3D analysis of these subsets no significant signal is present. This is not surprising given the small number of sources in the spectroscopic subsample.

\section{Discussion}\label{sec:discussion}

    \subsection{Robustness of the results}\label{fidelity}
\begin{figure}[t] 
    \centering
    \includegraphics[width=\columnwidth]{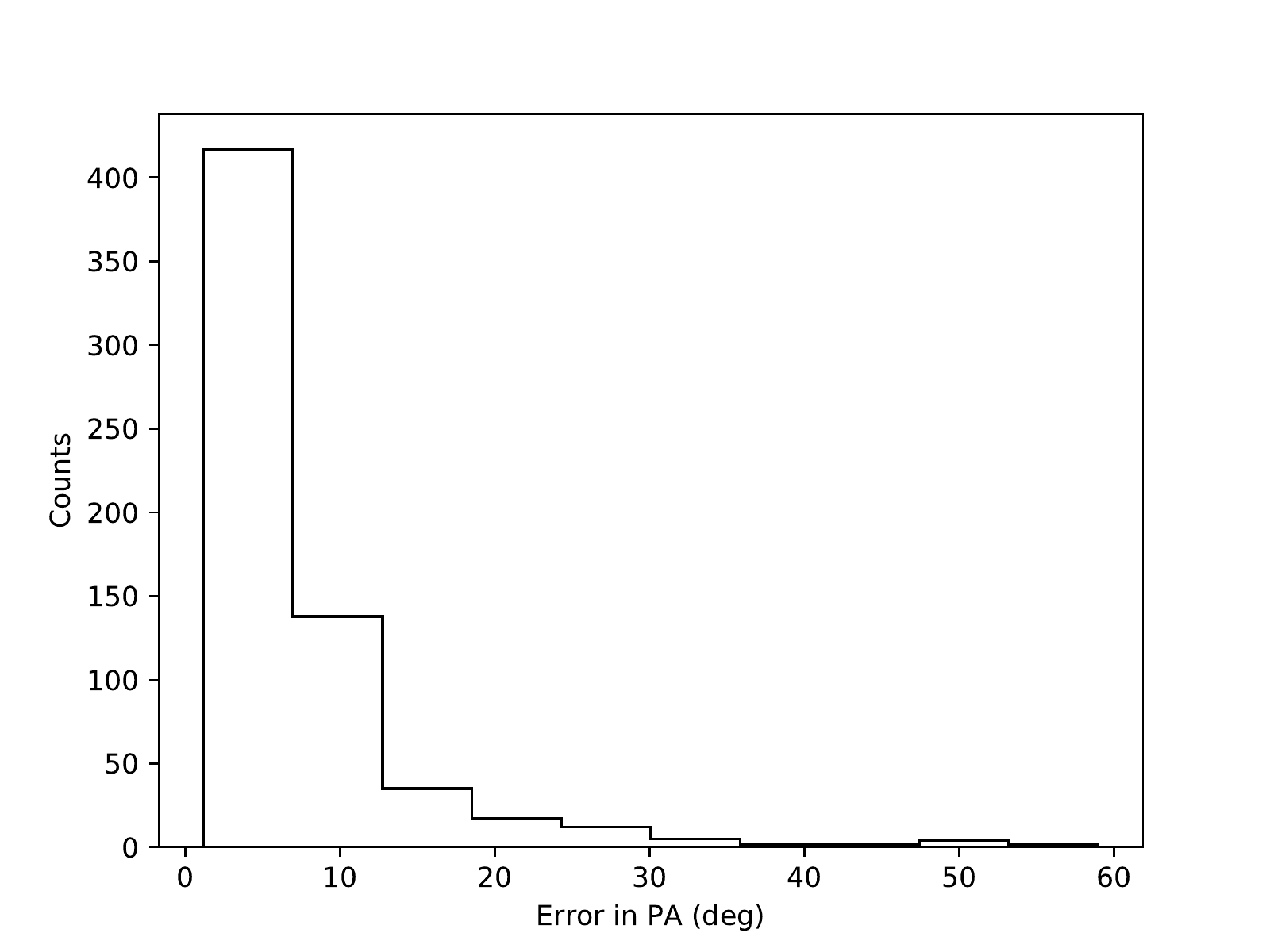}
    \caption{1$\sigma$ uncertainties on the position angles of sources in our sample that are not classified by the LOFAR galaxy zoo, but by the source finder only.}
    \label{fig:E_PA}
    \end{figure}

    The robustness of the results depends on the uncertainties of the position angles that were fit to the sources. 
    The position angles of sources in the LOFAR value-added source catalog do not all include uncertainties. The subset of sources that was classified in the LGZ project do not come with position angle uncertainties. To examine the position angle uncertainties, we are thus restricted to using the sources that are classified by the source finder PyBDSF only. 
    To examine the position angle uncertainties, we plot the $1\sigma$ uncertainties as given by the catalog for these sources. These are shown in Figure \ref{fig:E_PA}. The figure shows that 81\% of the sources have position angle uncertainties smaller than ten degrees.
    Thus, we can approximate the uncertainty in the final significance level by assuming every source in our sample has a $1\sigma$ uncertainty of $10.$  Since most sources have smaller uncertainties, this assumption is likely to overestimate the uncertainty in the fitted position angles and, thus, in the final significance level.
    
    To approximate the error in the final significance level as a function of a 1$\sigma$ error of ten degrees in the position angle, we must propagate this error through the statistical analysis of Section \ref{sec:statistics}. However, there is no straightforward procedure to define the general error on the extracted significance level as a function of the error on the measured position angles. Simple error propagation can be applied to the calculation of $S_{n}$, but it becomes complicated when a one-tailed significance level is extracted. This is due to the dependence of the significance level on the position of $S_{n}$ in the distribution of $S_{n}\vert_{MC}$ (Equation \ref{eq:SignificanceLevel}). If $S_{n}$ lies far from the mean of the normal distribution, a given change in $S_{n}$ will lead to a smaller change in significance level than when $S_{n}$ lies near the mean of the distribution of $S_{n}\vert_{MC}$. This is a direct effect of the cumulative normal distribution function being steepest near the mean and flattest near the edges.
    Moreover, considering that for every sample, $S_{n}\vert_{MC}$ is found by simulating 1,000 random datasets by randomly shuffling the position angles of the sources, the distribution of $S_{n}\vert_{MC}$ will be unique for every sample that we have considered. Therefore, we can only approximate the error on concrete results and cannot give a general $1\sigma$ confidence level that will apply for a range of samples.

    \begin{figure}[t]  
    \centering
    \includegraphics[width=\columnwidth]{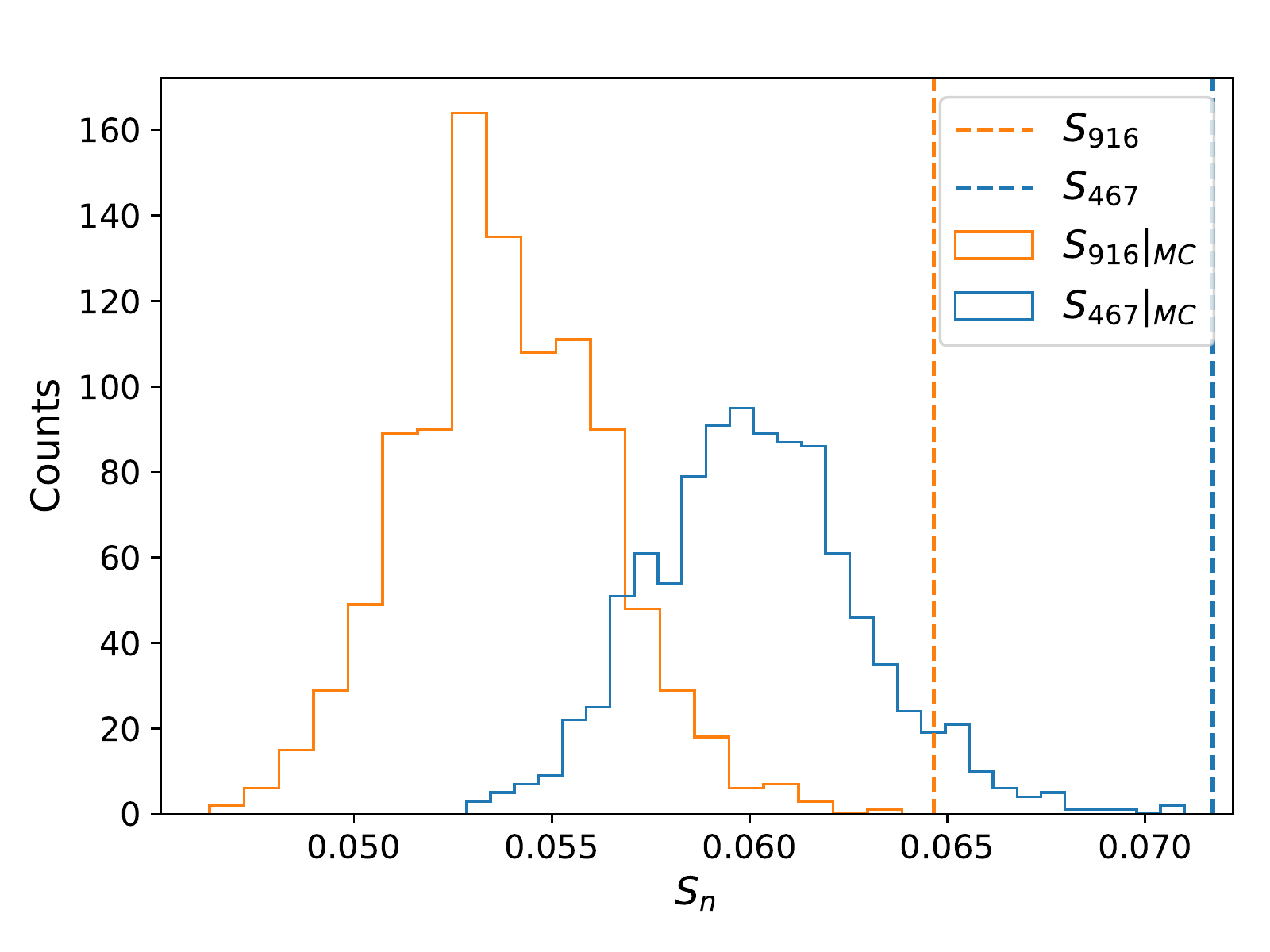}
    \caption{Distribution of the 1,000 simulated values $S_{n}\vert_{MC}$ and the highly significant values $S_{n}$ for $n = 467$ and $n = 916$. Plotted for the initial sample of sources.}
    \label{fig:s467_916initialsample}
    \end{figure}%
    
    The initial sample of sources rejected uniformity at a significance level of < $10^{-5}$ (Figure \ref{fig:SL_allsubsample}). The signal was found with a number of nearest neighbors between $467$ and $916$, corresponding to an angular scale between 3.1 and 4.6 degrees.
    Figure \ref{fig:s467_916initialsample} shows the distribution of the simulated data and the highly significant value of $S_{n}$ for these two bounds. We calculate the error on $S_{467}$ and $S_{916}$ and translate these errors to bounds on the significance values. 
    
    Assigning for each position angle in our sample a $1\sigma$ error of 10 degrees and applying standard error propagation, we find for the resulting values of $S_{467}$ and $S_{916}$, $0.070 \pm 0.0025$ and $ 0.065\pm 0.0026,$ respectively. Taking the 1$\sigma$ lower and upper bound of $S_{467}$ and calculating the significance level of these two bounds results in the lower and upper bound logarithmic significance levels of $-3.37$ and $-7.03$. For $S_{916}$ , the same method leads to lower and upper bound logarithmic significance levels of $-3.24$ and $-7.16$. Thus, strong evidence to reject uniformity in this sample at scales between 3.1 and 4.5 degrees is still found after applying possible uncertainties in the position angles. We can conclude that assuming a $1\sigma$ error of ten degrees on the position angle of all sources, the effect of an uncertainty in the position angles is quite powerful, but the significance level does remain strong enough to reject uniformity. Thus, the resulting significance level of $\log SL = -5.0$ to reject uniformity found in Figure \ref{fig:SL_allsubsample} for angular scales between 3.1 and 4.6 degrees should be stated with the approximate bound of $\log SL = -5.0 \pm 2.0$.

    As stated previously in this paper, the difference in significance level for the same variation in $S_{n}$ is dependent on the position of $S_{n}$, thus, it is also dependent on the significance level itself. Therefore, we reiterate that the change of two orders of magnitude in significance, found for the subset considered in this section, should not be applied to different subsets. We can apply the same calculation to the 3D analysis of the initial sample (Figure \ref{fig:SL_allsubsample3D}), where no result was found. We chose to investigate $n=500$, which corresponds to a significance level of $10^{-1.6}$. 
    This results in $\log SL = -1.7_{-0.97}^{+0.71}$; still without changing the signal to strong ($<10^{-3}$) evidence for alignment.
    Repeating the same calculations for the 2D analysis of the highest flux density sources that have an available redshift (Fig. \ref{fig:SL_flux3_2D_3D}) results in the approximate bounds $ \log SL = -5.3^{+1.4}_{-1.6}$ for $n$ = 100.

    \begin{figure}[t] 
    \centering
    \includegraphics[width=\columnwidth]{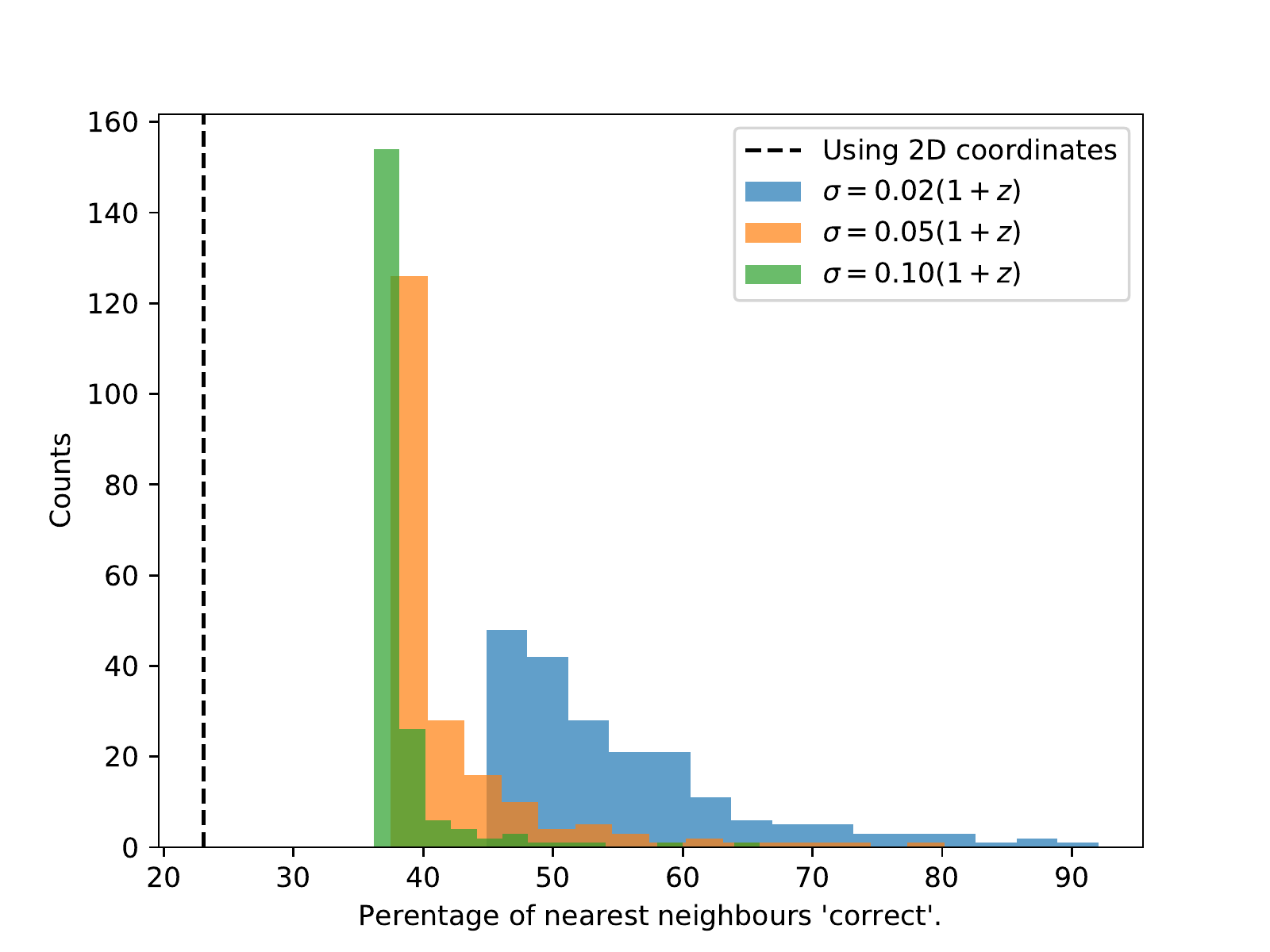}
    \caption{Percentage of the $n=101$ nearest neighbors of every source in the photo-z perturbed sample of 1,051 brightest sources that agrees with the nearest neighbors found in the unperturbed sample. See text for more details.}
    \label{fig:true_neighbours}
    \end{figure}%

    \begin{figure*}[t] 
    \centering
    \includegraphics[width=1.0\textwidth]{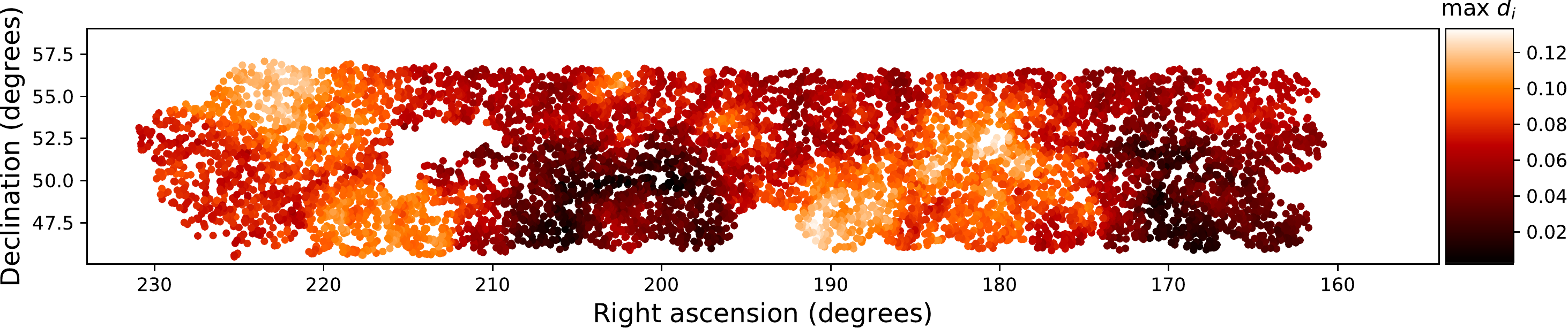}
    \caption{Scatter plot of the maximum dispersion measure (Equation \ref{eq:dispersionmax}) for every source, which indicates the strength of an alignment signal, of the selected sample of 7,555 radio sources plotted for $n=700$ as a function of right ascension and declination.}
    \label{fig:dispersion_RADEC_n700}
    \end{figure*}%

    \subsection{Interpretation of the results}\label{interpretation}

    Our complete sample of 7,555 double sources with a well-defined orientation was found to be inconsistent with a uniform distribution with a K-S test significance of 0.030 percent, which already indicates a global systematic effect in the data. However, the analysis of local alignment depends on the contrast between the statistic, $S_{n}$ , found for our dataset and the statistic, $S_{n}\vert_{MC}$ , found in absence of alignment. The statistic in absence of alignment was generated by randomly shuffling the position angles amongst the sources to maintain the same geometry and global position angle distribution. The advantage of this method over generating position angles from the uniform distribution $\mathcal{U}[0,180)$ is that it diminishes the effect of a possible global systematic present in our data sample. This is due to a global systematic then also being included in the distribution of the statistic $S_{n}\vert_{MC}$. Therefore, as long as $n \ll N$, where $N$ is the number of sources in the sample that is examined, the effect of the deviation from uniformity of the whole sample will not have considerably impacted the result of the significance of local alignment.
    
    To identify which particular sources are causing the observed signal, we examine which sources show the strongest alignment effect in 2D space. For this, we use the calculated maximum dispersion measure $d_{i,n}\vert_{max}$ (Equation \ref{eq:dispersionmax}), which measures the significance of the alignment of a source, $i,$ and its nearest neighbors, $n$ . We plot the maximum dispersion for every source in the initial sample of 7,555 sources as a function of right ascension and declination for $n = 700$ in Figure \ref{fig:dispersion_RADEC_n700}. From this figure, it becomes apparent that there is not a single region where the alignment is most pronounced, but rather, that there is an alternation between strongly aligned and less strongly aligned regions. This contradicts the observed effect being attributed to a survey-wide systematic effect, as then all sources would have similar maximum dispersion, regardless of their position. Additionally,  the scale of the alternation between aligned and non-aligned regions is larger than the typical separation between LOFAR pointings \citep[2.58 degrees;][]{2019A&A...622A...1S}, which makes the origin of the systematic effect even more elusive.
    
    We also found that the alignment signal was most significant for sources with the largest flux densities, as indicated by Figures \ref{fig:fluxbins2D} and \ref{fig:SL_flux3_2D_3D}. However, an analysis of the sources with the highest radio power did not show an alignment effect, either in 2D or in 3D. 
    Thus, it seems that only apparent source properties, rather than physical source properties, are correlated with the alignment effect, which could point towards an intrinsic effect of the survey, although radio power and source brightness are not strongly correlated for radio sources. Most importantly, the fact that the alignment effect is not present when using the 3D positions of the high flux density sources to find the nearest neighbors but is present when using 2D source positions (Fig. \ref{fig:SL_flux3_2D_3D}) may indicate a systematic error in the survey images or overall catalog, which is most noticeable or perhaps only present for the highest flux density sources. However, interpreting this result is not straightforward due to the relatively large uncertainties in the third (redshift) dimension.

    To further examine the impact of redshift uncertainties, we investigated whether 2D source positions are a better indicator of physical proximity than 3D source positions given different uncertainties in the photo-z estimates. This was done for the sample in Fig. \ref{fig:SL_flux3_2D_3D} with the 1,051 highest flux density sources that showed a signal in 2D around $n=101$ and no signal in 3D. We assumed, for this simulation, that the "true" source positions are given by the spectroscopic redshifts and best available photo-z estimates (i.e., that the photo-z scatters around the true redshift). The goal is to investigate what fraction of nearest neighbors that are found by using 3D positions agrees with the nearest neighbors found using the "true" source positions.
    
    The redshift of sources with a photo-z estimate was perturbed by a Gaussian with standard deviations of the usual form $\sigma(1+z)$ and spectroscopic redshifts were left intact. The $n=101$  nearest neighbors in 3D were then found for every source given the perturbed redshifts and the fraction of "correct" nearest neighbors was calculated. What we mean by "correct" here is that a nearest neighbor that was found is also one of the $n=100$ nearest neighbors using the "true" source positions, thus, we do not take the ordering into account (as the statistical method for a single value of $n$ does not do either). Figure \ref{fig:true_neighbours} shows the result of re-sampling the photometric redshifts 200 times and computing the fraction of "correct" nearest neighbors. The figure shows that using 2D coordinates leads to finding $23\%$ of the true physically close sources, while using 3D coordinates leads to finding more than $35\%$ of the physically close sources, even with standard deviations as as large as $0.1\times(1+z)$. Our assumed scatter of $0.1\times(1+z)$ represents a conservative upper limit on the expected precision of the LoTSS photometric redshift estimates, with the typical scatter for the radio population found to range from $0.03\times(1+z)$ for radio sources dominated by stellar emission and 0.08 to $0.1\times(1+z)$ for the more difficult quasar and AGN population \citep[see][]{2019A&A...622A...3D}.
    
    Thus, for this sample of sources and $n=101$, it would be likely to find a stronger alignment using 3D coordinates if the alignment effect is correlated with physical source positions. However, we are finding stronger alignment using 2D coordinates, which qualitatively implies that the alignment effect is more correlated with observed 2D source positions than it is with 3D source positions.

    \subsection{Scale of the alignment}
    
    The angular scale of the observed alignment effect is substantially larger than that of the two previous radio structure studies.
    \citet{2016MNRAS.459L..36T} investigated an area of 1.2 square degrees, and were thus limited to finding alignment within this area. Therefore, the angular scale of one degree found in that study might be underestimated and may still be in agreement with the results of this study. \citet{2017MNRAS.472..636C}, however, did not suffer this limitation, as they studied an area of 7000 square degrees and found an effect up to scales smaller than 2.5 degrees, with the maximum alignment signal at 1.5 degrees, while the distribution of source redshifts is not significantly different from that in this study. While the scale of the maximum effect does not agree with the angular scale of larger than three degrees found in this study, \citet{2017MNRAS.472..636C} limited their search to angular scales below 2.5 degrees, so the signal may perhaps be present on larger scales in the FIRST survey as well. Further research into radio jet alignment at larger angular scales is thus needed.
    
     Should the effect turn out to be physical, it is useful to compute the approximate physical scale corresponding to the effect that is observed. We computed the physical scale corresponding to the angular scale at which the alignment was found in this study by assuming the median redshift of the sample of sources for which a redshift is available (z = 0.56). Converting the angular scale of four degrees to comoving distances yields a corresponding physical scale of 103 $h^{-1}$Mpc. Although, as expected due to the limits in angular scales of the previous studies, the physical scale of 100 $h^{-1}$Mpc found in this study does not agree with the physical scale of the two previous studies of radio lobe alignment discussed earlier, it is in agreement with physical scales where other studies have found AGN alignment effects. As stated in Section 1, several studies have found that the radio polarization of quasars is preferentially aligned either perpendicular or parallel to the major axis of the surrounding large-scale large quasar groups (LQGs). These effects range from distances of the order of 150 Mpc \citep{2013IJMPD..2250089T} to distances larger than 300 $h^{-1}$Mpc \citep{2016A&A...590A..53P}. The physical scales found in this study agree with the physical scale for alignment with large-scale structures and coincides with the observed first peak of the Baryon Acoustic Oscillations \citep[BAO;][]{2005ApJ...633..560E}, while still abiding by the upper limits of homogeneity of the Universe, found to be on the order of 260 $h^{-1}$Mpc \citep[e.g.,][]{2010MNRAS.405.2009Y}.

\section{Conclusion}\label{sec:conclusion}

    In this study, we  analyze the uniformity in the position angles of extended radio sources with well-defined linear double structures from the initial installment of the LOFAR Two-metre Sky Survey (LoTSS). The combination of low frequencies (with sensitivity to extended structures) and the relatively high angular resolution of LOFAR makes it an excellent survey in the search for systematic alignments in the position angles of radio sources. 
    
    We extracted 7,555 LoTSS-extended sources with well-defined position angles from the 318,520 sources in the radio/optical value-added catalog of LOFAR sources in the HETDEX Spring Field region. To test for the alignment of position angles in this sample, the spherical nature of position angles and the effect of transporting these angles over the celestial sphere were taken into account using statistical methods originally developed to test for the alignment of polarization vectors. We find evidence for alignment in our initial sample of sources. The null hypothesis that the position angles are distributed uniformly can be rejected with a significance level of $<10^{-5}$ for an angular scale of four degrees, with the most non-uniformity present for radio sources with the largest flux densities. 
    
    Approximately half of the sources in our final sample have estimated redshifts available, either photometric or spectroscopic. This allows us to analyze the uniformity of radio source position angles in 3D space, but no strongly significant deviation from uniformity was found. We think it is more likely that the effect is caused by systematic effects, given the fact that the 2D analysis of the same reduced sample of sources still show an effect. However, the results are not straightforward to interpret due to the added uncertainties on the photometric redshifts, leaving no indisputable conclusion.

    Understanding the systematic effect or physical effect that causes the observed alignment in different radio surveys is beyond the scope of this study, but should be investigated further. In particular, these subtle effects will be important for cosmological analyses with radio data, such as weak lensing studies with the Square Kilometer Array \citep[e.g.,][]{2016MNRAS.463.3674H,2016MNRAS.463.3686B}. 

    The number of sources considered here comprises less than 2\% of the complete LoTSS survey. Hence, future studies by LOFAR should result in information about radio source alignments caused by substantially more subtle effects than we are presently able to determine. Additionally, the WEAVE-LOFAR project \citep{2016sf2a.conf..271S} will obtain over a million spectra of radio sources in LoTSS, which will allow for a much more detailed study of alignment in 3D space. This will provide the statistics needed to prove or disprove whether the alignment effect observed in this study is physical.

  \paragraph{}

\begin{acknowledgements}
      We kindly thank the anonymous referee for the comments and instructive insights. 
      EO and RJvW acknowledge support from the VIDI research programme with project number 639.042.729, which is financed by the Netherlands Organisation for Scientific Research (NWO). MJH acknowledges support from STFC [ST/R000905/1]. WLW acknowledges support from the ERC Advanced Investigator programme NewClusters 321271. WLW also acknowledges support from the CAS-NWO programme for radio astronomy with project number 629.001.024, which is financed by the Netherlands Organisation for Scientific Research (NWO). KJD acknowledges support from the ERC Advanced Investigator programme NewClusters 321271. APM would like to acknowledge the support from the NWO/DOME/IBM programme ``Big Bang Big Data: Innovating ICT as a Driver For Astronomy'', project \#628.002.001. HR acknowledges support from the ERC Advanced Investigator programme NewClusters 321271.
      LOFAR is the Low Frequency Array designed and constructed by ASTRON. It has observing, data processing, and data storage facilities in several countries, which are owned by various parties (each with their own funding sources), and which are collectively operated by the ILT foundation under a joint scientific policy. The ILT resources have benefitted from the following recent major funding sources: CNRS-INSU, Observatoire de Paris and Université d'Orléans, France; BMBF, MIWF-NRW, MPG, Germany; Science Foundation Ireland (SFI), Department of Business, Enterprise and Innovation (DBEI), Ireland; NWO, The Netherlands; The Science and Technology Facilities Council, UK; Ministry of Science and Higher Education, Poland.

      This research made use of the Dutch national e-infrastructure with support of the SURF Cooperative (e-infra 180169) and the LOFAR e-infra group. The Jülich LOFAR Long Term Archive and the German LOFAR network are both coordinated and operated by the Jülich Supercomputing Centre (JSC), and computing resources on the Supercomputer JUWELS at JSC were provided by the Gauss Centre for Supercomputing e.V. (grant CHTB00) through the John von Neumann Institute for Computing (NIC).

      This research made use of the University of Hertfordshire high-performance computing facility and the LOFAR-UK computing facility located at the University of Hertfordshire and supported by STFC [ST/P000096/1].
\end{acknowledgements}

\bibliographystyle{aa}
\bibliography{firstbib}

\end{document}